\documentclass{PoS}
\usepackage{amsmath}
\usepackage{xspace}

\newcommand{\MCatNLO}{M\protect\scalebox{0.8}{C}@N\protect\scalebox{0.8}{LO}\xspace}
\newcommand{\POWHEG}{P\protect\scalebox{0.8}{OWHEG}\xspace}
\newcommand{\MEPS}{M\scalebox{0.8}{E}P\scalebox{0.8}{S}\xspace}
\newcommand{\NLOPS}{N\scalebox{0.8}{LO}P\scalebox{0.8}{S}\xspace}
\newcommand{\MENLOPS}{ME\protect\scalebox{0.8}{NLO}PS\xspace}
\newcommand{\MEPSatNLO}{M\scalebox{0.8}{E}P\scalebox{0.8}{S}@N\protect\scalebox{0.8}{LO}\xspace}

\newcommand{\Sherpa}{S\protect\scalebox{0.8}{HERPA}\xspace}

\newcommand{\done}{\mathrm{d}}

\newcommand{\mr}{\mathrm}

\newcommand{\abr}[1]{\langle #1\rangle}

\pdfoutput=1

\title{\vspace{-3.5cm}{\normalsize IPPP/12/91, DCPT/12/182, LPN12-131, SLAC--PUB--15305}\\\vspace{2cm}
       Systematic uncertainties in \protect\NLOPS matching}
\ShortTitle{Systematic uncertainties in \protect\NLOPS matching}

\author{\speaker{Marek Sch\"onherr}\footnote{Supported by the by the Research 
        Executive Agency (REA) of the European Union under the Grant Agreement 
        number PITN-GA-2010-264564 (LHCPhenoNet).}\\
        Institute for Particle Physics Phenomenology,
        Durham University, Durham DH1 3LE, UK\\
        E-mail: \email{marek.schoenherr@durham.ac.uk}}

\author{Stefan H\"oche\\
        SLAC National Accelerator Laboratory, 
        Menlo Park, CA 94025, USA\\\vspace*{-4mm}}

\author{Frank Krauss\\
        Institute for Particle Physics Phenomenology,
        Durham University, Durham DH1 3LE, UK\\\vspace*{-4mm}}

\author{Frank Siegert\\
        Physikalisches Institut,
        Albert-Ludwigs-Universit{\"a}t Freiburg, D-79104 Freiburg, Germany\\\vspace*{-4mm}}

\abstract{
	  The \MCatNLO and \MEPSatNLO methods, as implemenated in the 
	  Monte-Carlo event generator framework \Sherpa, are used to 
	  estimate the perturbative and non-perturbative uncertainties 
	  in various processes such as dijet production and the production 
	  of a $W$ boson in association with (multiple) jets.
	 }

\FullConference{36th International Conference on High Energy Physics,\\
		July 4-11, 2012\\
		Melbourne, Australia}

\begin{document}

\section{Introduction}

Being largely stimulated by the need for higher precission of theoretical 
predictions in both Standard Model analyses and new physics searches at the 
LHC, the simulation of higher-order QCD corrections in Monte Carlo event 
generators has seen vast improvements in recent years. To this end, 
two lines of development have been followed. In the \MEPS approach 
\cite{Catani:2001cc,Lonnblad:2001iq,Krauss:2002up,Hoeche:2009rj,
  Hamilton:2009ne,LonnBlad:2011xx}
higher-order tree-level matrix elements of successive final state parton 
multiplicity are merged into an inclusive sample, offering both leading-order 
accuracy for the production of hard partons and retaining the overall 
resummation of scale hierarchies through the parton shower at the same 
time. On the other hand \NLOPS approaches, introduced as either \MCatNLO 
\cite{Frixione:2002ik} or \POWHEG \cite{Nason:2004rx,Frixione:2007vw}, work 
on a single parton multiplicity elevating its accuracy to next-to-leading 
order. Both methods have been shown to be automatable 
\cite{Hoeche:2010pf,Hoeche:2011fd} within the \Sherpa event generator 
framework \cite{Gleisberg:2008ta}. Thereafter, it was sought to recombine both 
lines of development. In a first step, called the \MENLOPS prescription, the 
\NLOPS and \MEPS 
methods have been combined using the \NLOPS' NLO accuracy for the inclusive 
process supplementing it with higher-order tree-level matrix elements in 
an \MEPS fashion \cite{Hoeche:2010kg}. In second step multiple \NLOPS 
processes of successive parton multiplicity are combined, elevating the 
accuracy of the \MEPS method to next-to-leading order, dubbed \MEPSatNLO 
\cite{Hoeche:2012yf,Gehrmann:2012yg}. In the following both the \NLOPS and 
\MEPSatNLO methods are summarised. Particular emphasis is put on both 
methods' major accompishments with respect to standard leading order 
computations: its increased theoretical accuracy expressed through reduced 
perturbative uncertainties.

\section{\protect\NLOPS matching}

Following the notation of \cite{Hoeche:2011fd} a general NLO+PS matching can 
be cast in the form of the following master formula
\begin{equation}
  \begin{split}
    \langle O\rangle 
    \,=&\,
    \int\done\Phi_B
      \;\;\bar{\rm B}^\text{(A)}(\Phi_B)\;\vphantom{\bigg|}
      \Bigg[\Delta^\text{(A)}(t_0,\mu_Q^2)\,O(\Phi_B)
	    +\sum_i\int_{t_0}^{\mu_Q^2}\done\Phi_1\,
	      \frac{{\rm D}_i^\text{(A)}(\Phi_B,\Phi_1)}
		  {{\rm B}(\Phi_B)}\,
	      \Delta^\text{(A)}(t,\mu_Q^2)\,O(\Phi_R)
      \Bigg]\\
    &\;{}
      +\int\done\Phi_R\; {\rm H}(\Phi_R)\;O(\Phi_R)\,.
  \end{split}
\end{equation}
Therein, the NLO-weighted normalisation of the resummed events is defined as
\begin{equation}
  \bar{\rm B}^\text{(A)}(\Phi_B)
  ={\rm B}(\Phi_B)+\tilde{\rm V}(\Phi_B)+{\rm I}^\text{(A)}(\Phi_B)
  +\sum_i\int\done\Phi_1\left[{\rm D}_i^\text{(A)}
  \Theta(\mu_Q^2-t)-{\rm D}_i^\text{(S)}\right](\Phi_B,\Phi_1)\,.
\end{equation}
$t=t(\Phi_1)$ identifies the infrared limits of the additional parton's phase 
space and serves as an ordering 
variable of the parton shower resummation. The resummation kernels are then 
defined by the auxiliary set of subtraction kernels ${\rm D}^\text{(A)}$, 
ensuring the correct behaviour in both the soft and the collinear limit of 
the emission of an extra parton, exhibiting full colour and spin correctness 
in the respective limits. They imply the modified Sudakov form factor
\begin{equation}
  \Delta^\text{(A)}(t_0,t_1)
  = \exp\left[-\int_{t_0}^{t_1}\!\!\!\!\done\Phi_1\,
		\frac{{\rm D}_i^\text{(A)}(\Phi_B,\Phi_1)}{{\rm B}(\Phi_B)}\right]\,.
\end{equation}

\begin{figure}[b]
  \includegraphics[width=0.47\textwidth]{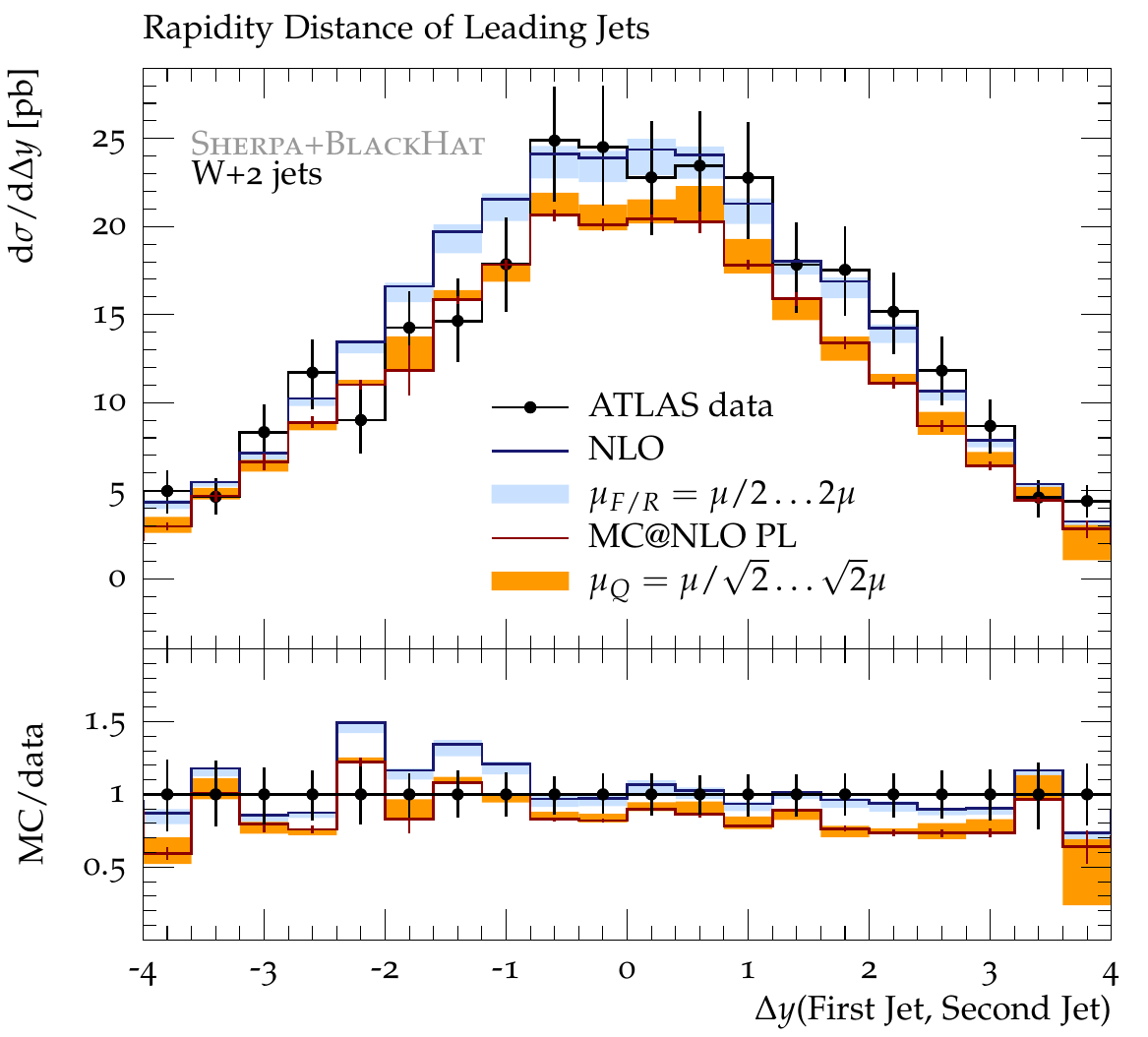}
  \hfill
  \includegraphics[width=0.47\textwidth]{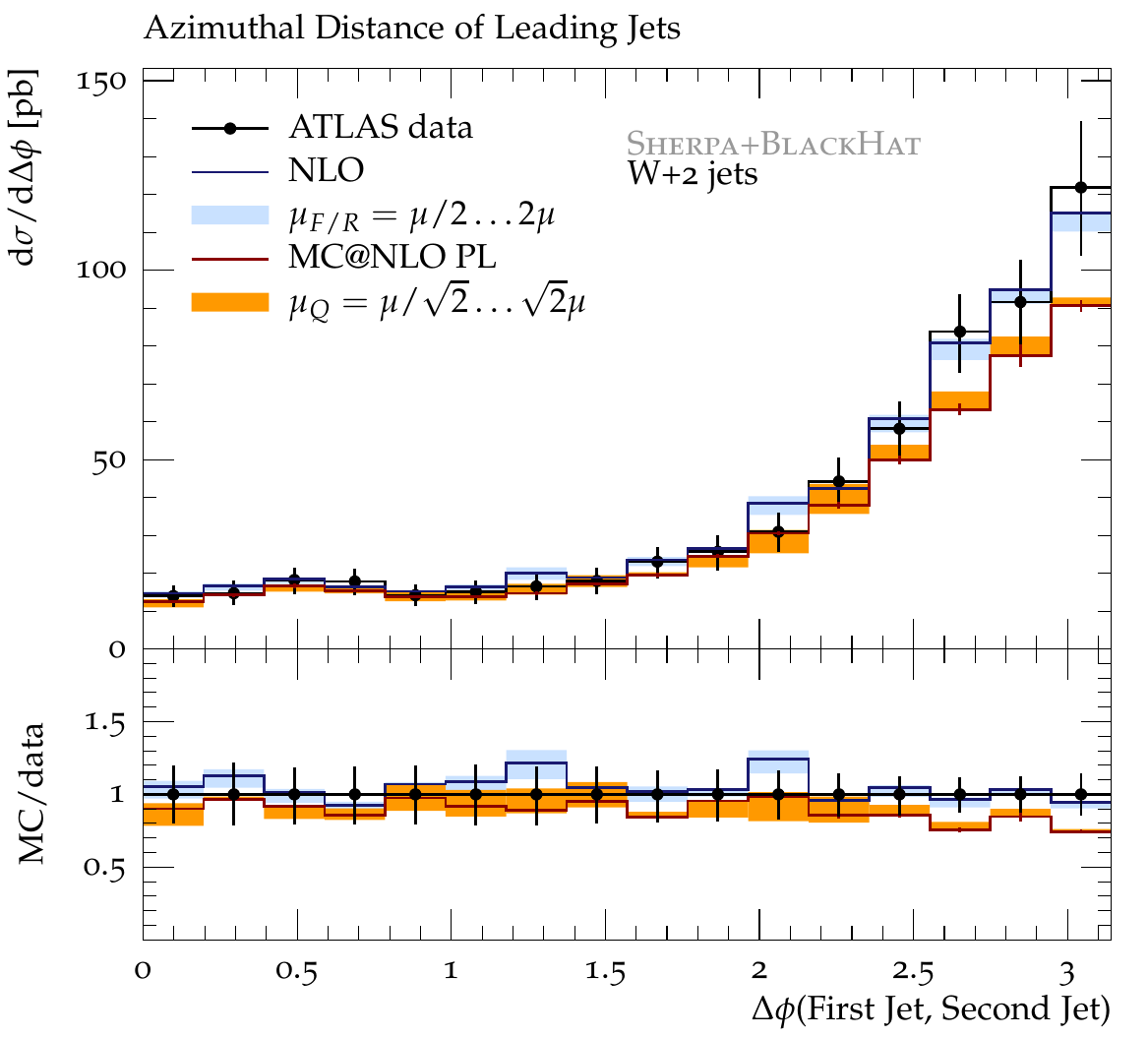}
  \caption{
	   Rapidity (left) and azimuthal (right) separation of the two leading 
	   jet $pp\to\ge 2\,\text{jets}$ compared to ATLAS data \cite{Aad:2012en}.
	   \label{Fig:wnj}
	  }
\end{figure}

\begin{figure}[b!]
  \includegraphics[width=0.47\textwidth]{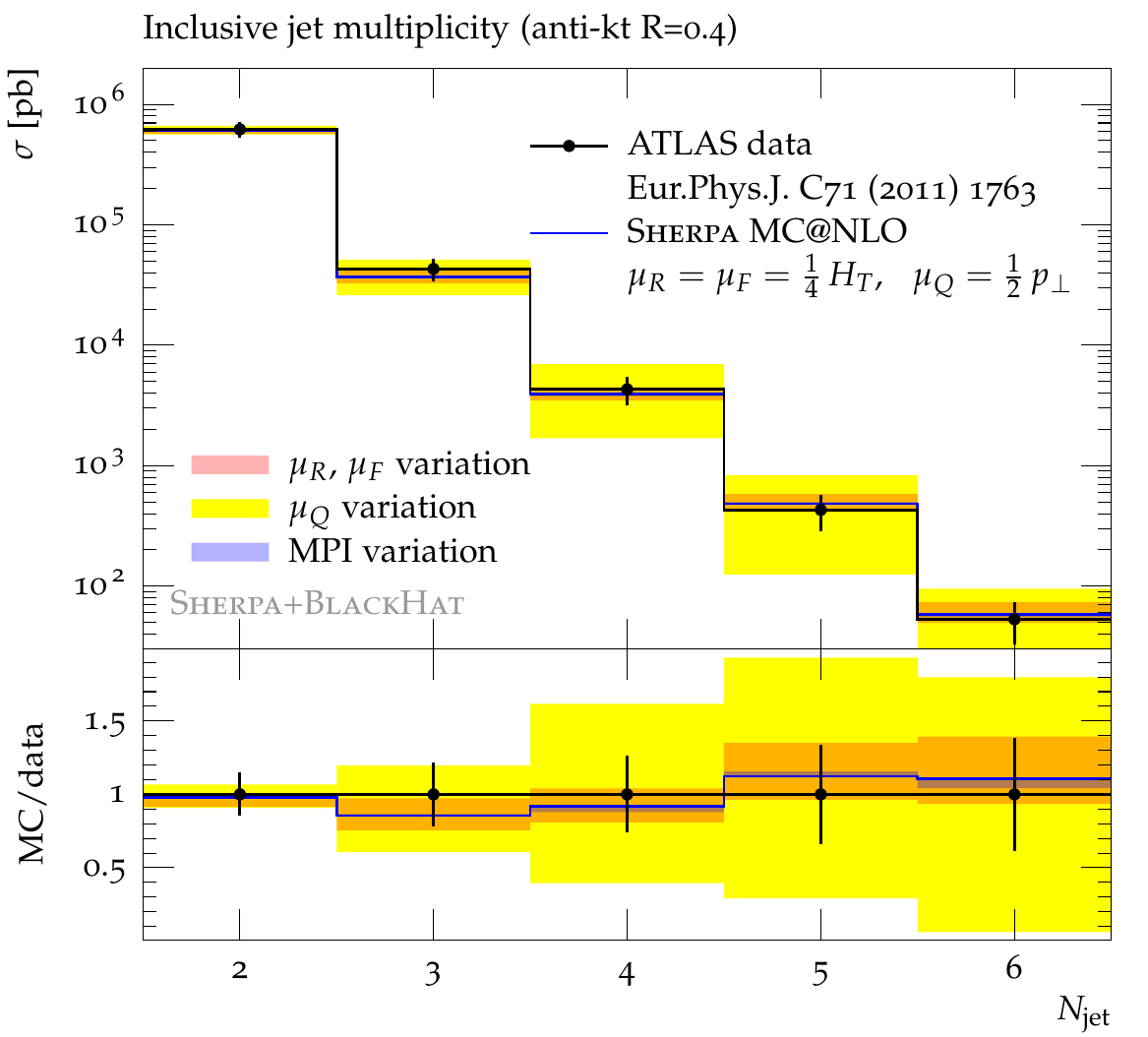}
  \hfill
  \includegraphics[width=0.47\textwidth]{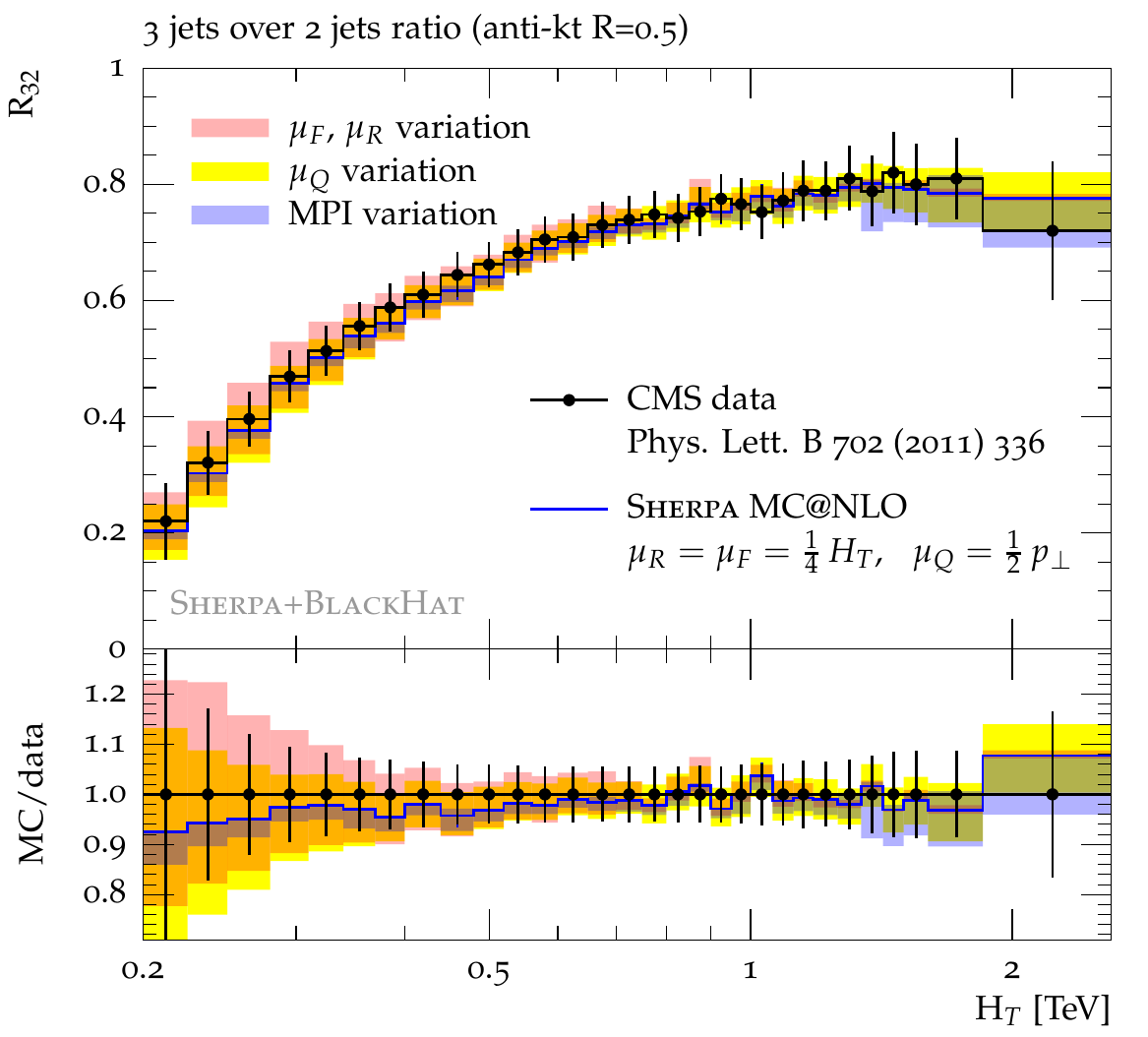}
  \caption{
           {\bf Left:} Inclusive jet cross section in $pp\to\ge 2\,\text{jets}$ 
           compared to ATLAS data \cite{Aad:2011tqa}. {\bf Right:} 3-jet over 
           2-jet ratio in dependence on the scalar transverse momentum sum of 
           all jets in $pp\to\ge 2\,\text{jets}$ in comparison to CMS 
           \cite{Chatrchyan:2011wn}.
           \label{Fig:ATLAS_incl_jet_rates_CMS_R32}
          }
\end{figure}

An upper scale $\mu_Q$ limits the region of resummation, i.e. the exponent of 
the Sudakov form factor vanishes at $t=\mu_Q$. This scale has been made 
accessible for the first time in the implementation of \cite{Hoeche:2011fd} 
and can thus be used to study the uncertaity related to its arbitrariness. 
The finite remainder of the real emission cross section is then embedded in 
the so-called hard events defined through
\begin{equation}
  {\rm H}(\Phi_R)
  ={\rm R}(\Phi_R)-\sum_i{\rm D}_i^\text{(A)}(\Phi_R)\,.
\end{equation}
Fig.\ \ref{Fig:wnj} now shows an evaluation of the resummation scale 
uncertainty in various \MCatNLO implementations for $pp\to W+n\,\text{jets}$
\cite{Hoeche:2012ft}
and contrasts it with the renormalisation and factorisation scale 
uncertainties in a standard fixed-order next-to-leading order calculation. 
Fig.\ \ref{Fig:ATLAS_incl_jet_rates_CMS_R32} details all sources of perturbative 
($\mu_R$, $\mu_F$, $\mu_Q$) as well as non-perturbative uncertainties due to 
the multiple interaction model in an \MCatNLO implementation of inclusive and 
dijet production \cite{Hoeche:2012fm}. In all cases, the perturbative 
uncertainties for observables described at NLO accuracy are greatly reduced 
while the parton shower resummation provides the correct description when 
large hierarchies of scales in $t$ are present. At the same time, there are 
observables/regions where the uncertainty on the modelling of the soft 
structure of the event is non-negligible.

\section{\protect\MEPSatNLO merging}

The \NLOPS matched calculations detailed in the previous 
section can now be used as input to extend the CKKW-type
to next-to-leading order \cite{Gehrmann:2012yg,Hoeche:2012yf}.
The master formula for its construction reads as follows
\begin{equation}\label{eq:nlomerging}
  \begin{split}
  \abr{O}\,=&\;
  \int\done\Phi_n\;\bar{\rm B}_n^\text{(A)}
    \Bigg[
	\Delta_n^\text{(A)}(t_c,\mu_Q^2)\,O_n
        +\int\limits_{t_c}^{\mu_Q^2}\done\Phi_1\;
         \frac{{\rm D}_n^\text{(A)}}{{\rm B}_n}\,
	 \Delta_n^\text{(A)}(t_{n+1},\mu_Q^2)\,\Theta(Q_\text{cut}-Q_{n+1})\;
         O_{n+1}\;
  \,\Bigg]\\
  &{}
    +\int\done\Phi_{n+1}\;\mr{H}_n^{\rm(A)}\,
    \Delta_n^\text{(PS)}(t_{n+1},\mu_Q^2)\,
   \Theta(Q_\text{cut}-Q_{n+1})\;O_{n+1}\\
  &{}
   +\int\done\Phi_{n+1}\;\bar{\rm B}_{n+1}^\text{(A)}
   \Bigg(\,
     1+\frac{{\rm B}_{n+1}}{\bar{\rm B}_{n+1}^\text{(A)}}
     \int\limits_{t_{n+1}}^{\mu_Q^2}\done\Phi_1\,{\rm K}_n\,
     \Bigg)
   \Delta_n^\text{(PS)}(t_{n+1},\mu_Q^2)\,\Theta(Q_{n+1}-Q_\text{cut})\\
  &\hspace*{20mm}
    \times\Bigg[
	\Delta_{n+1}^\text{(A)}(t_c,t_{n+1})\,O_{n+1}
        +\int\limits_{t_c}^{t_{n+1}}\done\Phi_1\;
         \frac{{\rm D}_{n+1}^\text{(A)}}{{\rm B}_{n+1}}\,
	 \Delta_{n+1}^\text{(A)}(t_{n+2},t_{n+1})\;
         O_{n+2}\;
  \,\Bigg]\\
  &{}
  +\int\done\Phi_{n+2}\;\mr{H}_{n+1}^{\rm(A)}\,
  \Delta_{n+1}^\text{(PS)}(t_{n+2},t_{n+1})\,\Delta_n^\text{(PS)}(t_{n+1},\mu_Q^2)\,
  \Theta(Q_{n+1}-Q_\text{cut})\;O_{n+2}\;+\,\ldots\;,
  \end{split}
\end{equation}
Therein an \MCatNLO description of an $n$ parton multiplicity is restricted 
to have its emission produced at a jet measure $Q$ smaller than $Q_\text{cut}$. 
The region with $Q>Q_\text{cut}$ is then filled with an \MCatNLO for the $n+1$ 
parton process. To restore the correct resummation with respect to the $n$ 
parton process to at least parton shower accuracy its Sudakov form factor 
$\Delta_n^\text{(PS)}$ is inserted. The overlap with similar terms in 
$\bar{\rm B}_{n+1}^\text{(A)}$ is removed with the term in the braces on third 
line. A multijet merged description is then achieved by iteration eq.\ 
\ref{eq:nlomerging}.

\begin{figure}[p]
  \centering
  \includegraphics[width=0.5\textwidth]{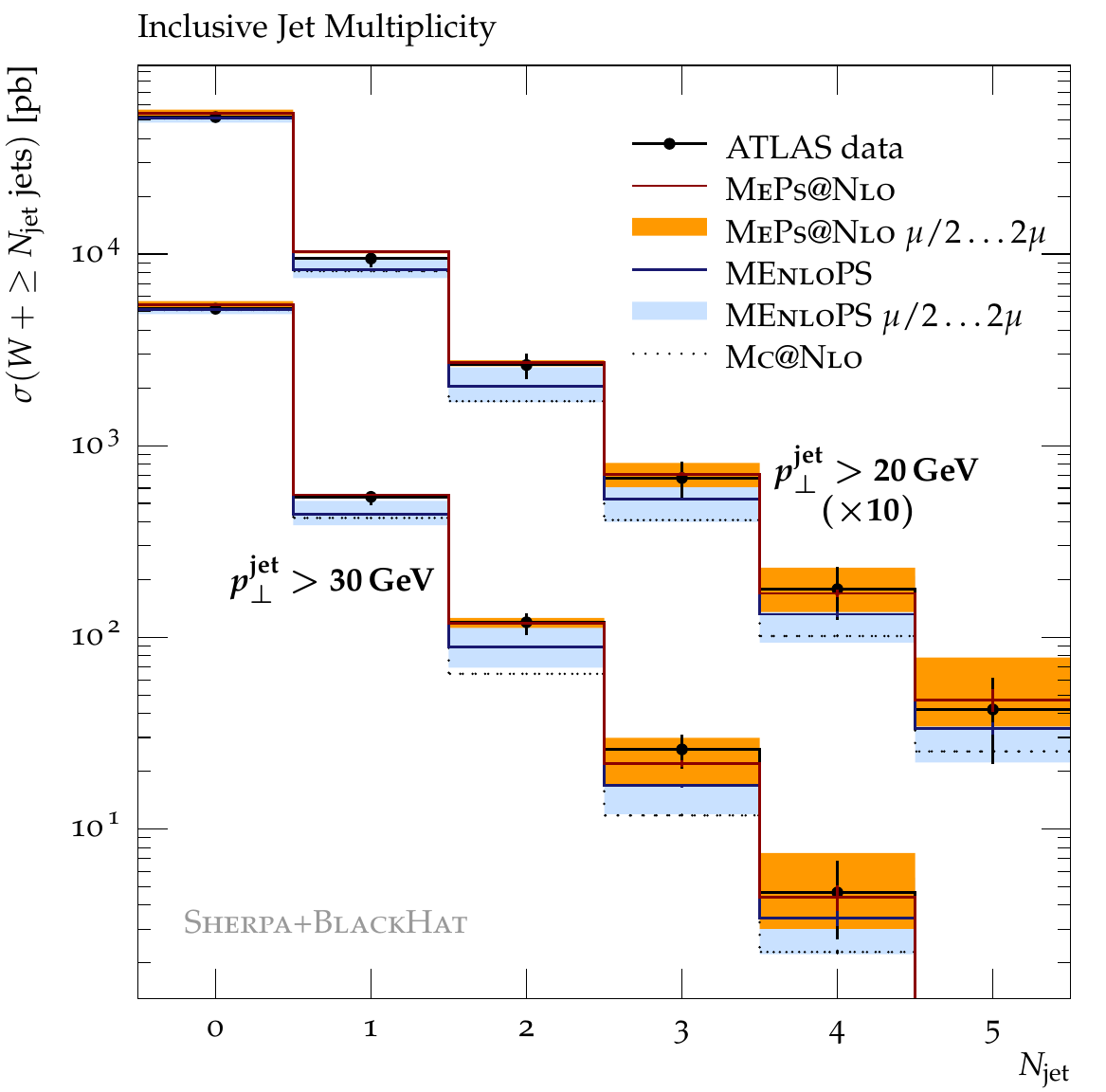}
  \caption{
          Cross section as a function of the inclusive jet multiplicity
          in $pp\to W+\text{jets}$ events compared to ATLAS data 
          \cite{Aad:2012en}.
          \label{Fig:ATLAS_jet_multi}
         }
\end{figure}

\begin{figure}[p]
  \begin{minipage}{0.47\textwidth}
    \lineskip-1.85pt
    \includegraphics[width=\textwidth]{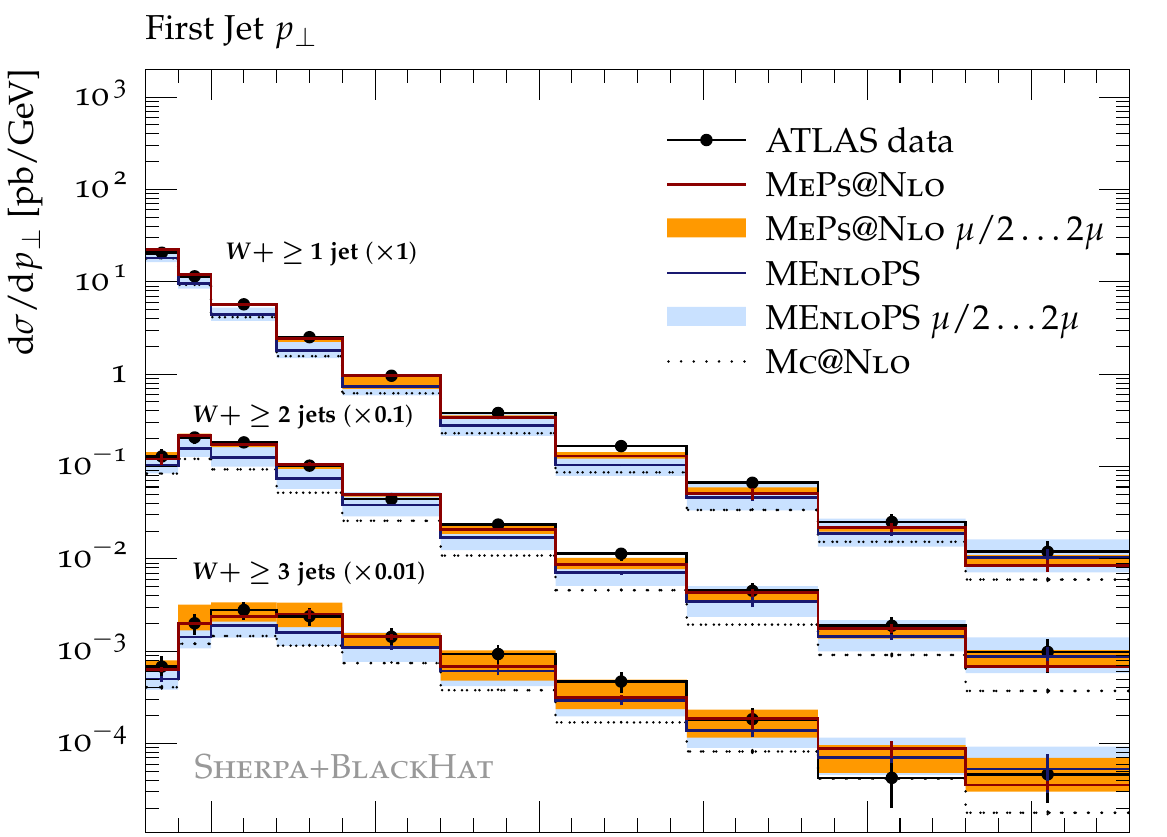}\\
    \includegraphics[width=\textwidth]{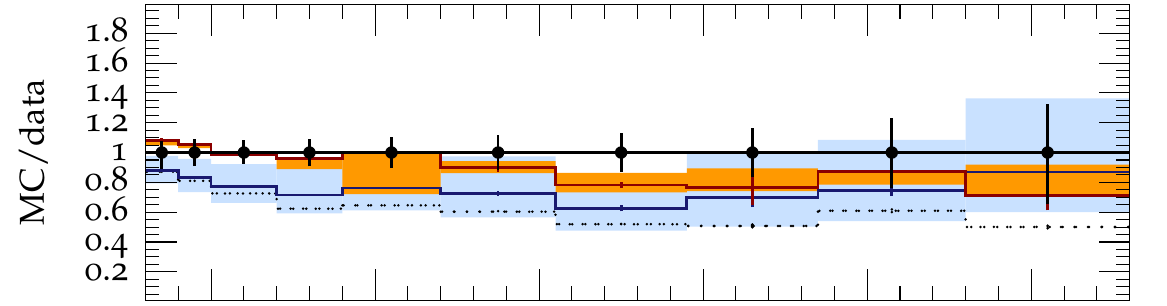}\\
    \includegraphics[width=\textwidth]{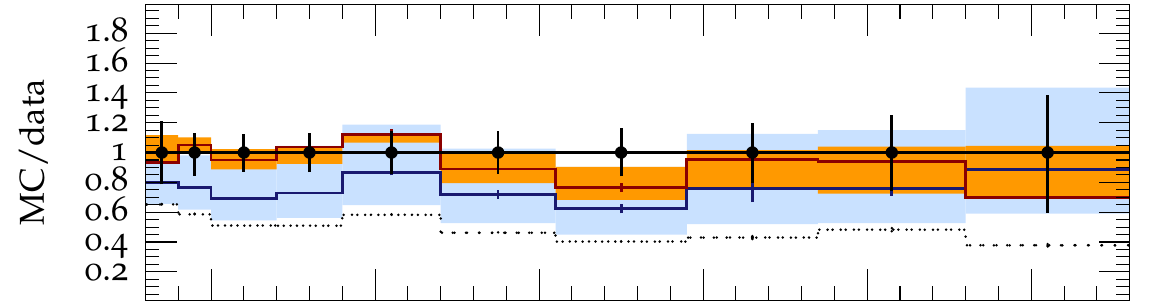}\\
    \includegraphics[width=\textwidth]{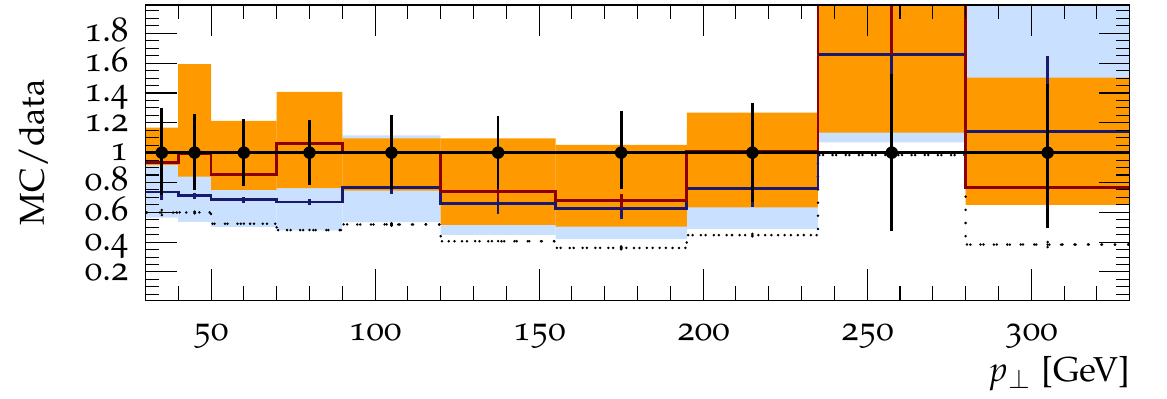}
  \end{minipage}\hfill
  \begin{minipage}{0.47\textwidth}
    \lineskip-1.85pt
    \includegraphics[width=\textwidth]{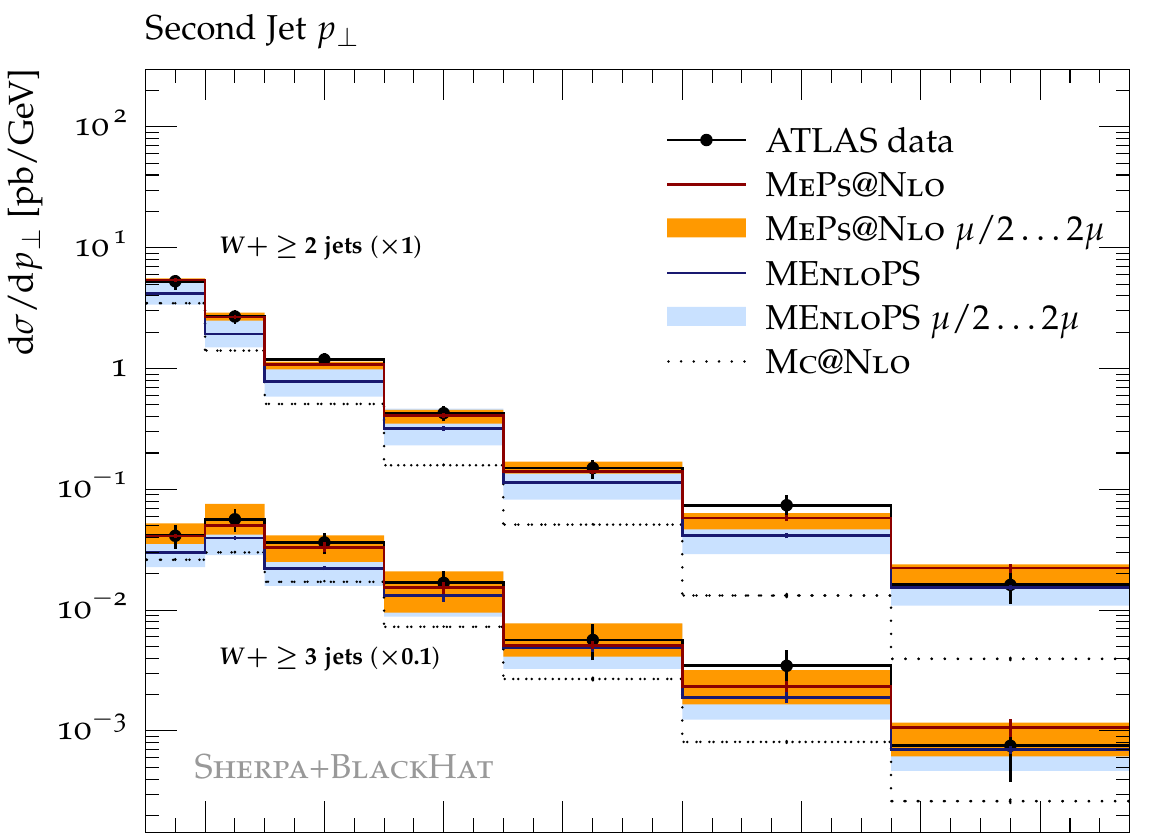}\\
    \includegraphics[width=\textwidth]{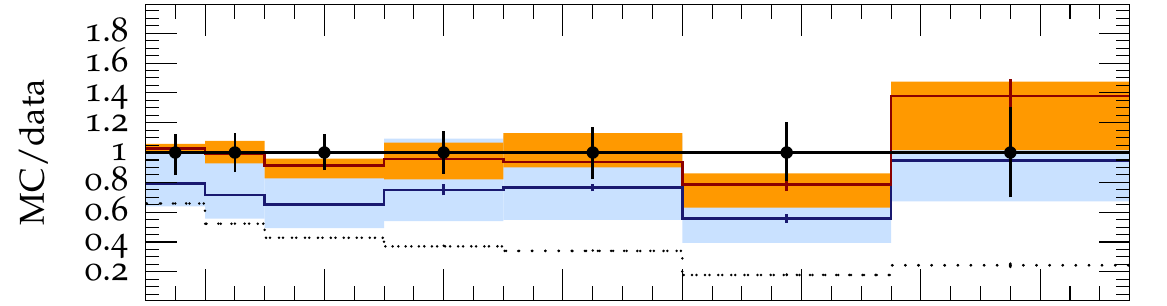}\\
    \includegraphics[width=\textwidth]{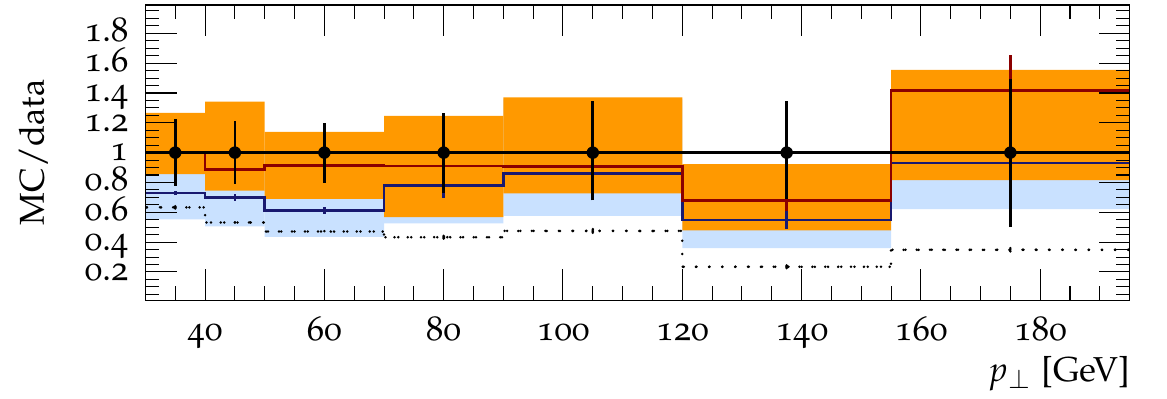}
    \vspace*{36pt}
  \end{minipage}
  \caption{
	   Differential cross section as a function of the transverse momentum 
	   of the first (left) and second (right) jet in 
	   $pp\to W+\ge 1,2,3\,\text{jets}$ events compared to ATLAS data 
	   \cite{Aad:2012en}.
	   \label{Fig:ATLAS_jetpts}
	  }
\end{figure}

Again, the calculation benefits from the decreased theoretical uncertainty 
of its \MCatNLO input processes. Figs.\ \ref{Fig:ATLAS_jet_multi} and 
\ref{Fig:ATLAS_jetpts} exemplify this feature for the process 
$pp\to W+\,\text{jets}$ compared to ATLAS data. For this calculation the 
processes with 0, 1 and 2 additional jets are described at next-to-leading 
order while 3 and 4 addiotional jets have been merged on top of that at leading 
order accuracy. These different levels of accuracy can be directly seen in the 
respective uncertainties. Further, they are contrasted with a \MENLOPS 
\cite{Hoeche:2010kg} prediction using an \MCatNLO input only for the $pp\to W$ 
process and merging only leading order prediction for 1, 2, 3 and 4 additional 
jets on top.

\bibliographystyle{amsunsrt_mod}
\bibliography{journal}

\begin{thebibliography}{10}

\bibitem{Catani:2001cc}
S.~Catani, F.~Krauss, R.~Kuhn and B.~R. Webber, \emph{{QCD matrix elements +
  parton showers}}, JHEP \textbf{11} (2001),
  \href{http://www.slac.stanford.edu/spires/find/hep/www?eprint=hep-ph/0109231%
}{063},  [\href{http://arXiv.org/pdf/hep-ph/0109231}{{\tt hep-ph/0109231}}].
  \relax
 \relax
\bibitem{Lonnblad:2001iq}
L.~L{\"o}nnblad, \emph{{Correcting the colour-dipole cascade model with fixed
  order matrix elements}}, JHEP \textbf{05} (2002),
  \href{http://www.slac.stanford.edu/spires/find/hep/www?eprint=hep-ph/0112284%
}{046},  [\href{http://arXiv.org/pdf/hep-ph/0112284}{{\tt hep-ph/0112284}}].
  \relax
 \relax
\bibitem{Krauss:2002up}
F.~Krauss, \emph{{Matrix elements and parton showers in hadronic
  interactions}}, JHEP \textbf{0208} (2002),
  \href{http://www.slac.stanford.edu/spires/find/hep/www?eprint=hep-ph/0205283%
}{015},  [\href{http://arXiv.org/pdf/hep-ph/0205283}{{\tt hep-ph/0205283}}].
  \relax
 \relax
\bibitem{Hoeche:2009rj}
S.~H{\"o}che, F.~Krauss, S.~Schumann and F.~Siegert, \emph{{QCD matrix elements
  and truncated showers}}, JHEP \textbf{05} (2009),
  \href{http://www.slac.stanford.edu/spires/find/hep/www?eprint=arXiv:0903.121%
9}{053},  [\href{http://arXiv.org/pdf/0903.1219}{{\tt arXiv:0903.1219}}
  [hep-ph]]. \relax
 \relax
\bibitem{Hamilton:2009ne}
K.~Hamilton, P.~Richardson and J.~Tully, \emph{{A modified CKKW matrix element
  merging approach to angular-ordered parton showers}}, JHEP \textbf{11}
  (2009),
  \href{http://www.slac.stanford.edu/spires/find/hep/www?eprint=arXiv:0905.307%
2}{038},  [\href{http://arXiv.org/pdf/0905.3072}{{\tt arXiv:0905.3072}}
  [hep-ph]]. \relax
 \relax
\bibitem{LonnBlad:2011xx}
L.~L{\"o}nnblad and S.~Prestel, \emph{{Matching Tree-Level Matrix Elements with
  Interleaved Showers}}, JHEP \textbf{03} (2012),
  \href{http://www.slac.stanford.edu/spires/find/hep/www?eprint=1109.4829}{019%
},  [\href{http://arXiv.org/pdf/1109.4829}{{\tt arXiv:1109.4829}} [hep-ph]].
  \relax
 \relax
\bibitem{Frixione:2002ik}
S.~Frixione and B.~R. Webber, \emph{{Matching NLO QCD computations and parton
  shower simulations}}, JHEP \textbf{06} (2002),
  \href{http://www.slac.stanford.edu/spires/find/hep/www?eprint=hep-ph/0204244%
}{029},  [\href{http://arXiv.org/pdf/hep-ph/0204244}{{\tt hep-ph/0204244}}].
  \relax
 \relax
\bibitem{Nason:2004rx}
P.~Nason, \emph{{A new method for combining NLO QCD with shower Monte Carlo
  algorithms}}, JHEP \textbf{11} (2004),
  \href{http://inspirebeta.net/record/659055}{040},
  [\href{http://arXiv.org/pdf/hep-ph/0409146}{{\tt hep-ph/0409146}}]. \relax
 \relax
\bibitem{Frixione:2007vw}
S.~Frixione, P.~Nason and C.~Oleari, \emph{{Matching NLO QCD computations with
  parton shower simulations: the POWHEG method}}, JHEP \textbf{11} (2007),
  \href{http://www.slac.stanford.edu/spires/find/hep/www?eprint=arXiv:0709.209%
2}{070},  [\href{http://arXiv.org/pdf/0709.2092}{{\tt arXiv:0709.2092}}
  [hep-ph]]. \relax
 \relax
\bibitem{Hoeche:2010pf}
S.~H{\"o}che, F.~Krauss, M.~Sch{\"o}nherr and F.~Siegert, \emph{{Automating the
  POWHEG method in \Sherpa}}, JHEP \textbf{04} (2011),
  \href{http://inspirebeta.net/record/866705}{024},
  [\href{http://arXiv.org/pdf/1008.5399}{{\tt arXiv:1008.5399}} [hep-ph]].
  \relax
 \relax
\bibitem{Hoeche:2011fd}
S.~H{\"o}che, F.~Krauss, M.~Sch{\"o}nherr and F.~Siegert, \emph{{A critical
  appraisal of NLO+PS matching methods}}, JHEP \textbf{09} (2012),
  \href{http://inspirehep.net/record/944643}{049},
  [\href{http://arXiv.org/pdf/1111.1220}{{\tt arXiv:1111.1220}} [hep-ph]].
  \relax
 \relax
\bibitem{Gleisberg:2008ta}
T.~Gleisberg, S.~H{\"o}che, F.~Krauss, M.~Sch\"{o}nherr, S.~Schumann,
  F.~Siegert and J.~Winter, \emph{{Event generation with \Sherpa 1.1}}, JHEP
  \textbf{02} (2009), \href{http://inspirebeta.net/record/803708}{007},
  [\href{http://arXiv.org/pdf/0811.4622}{{\tt arXiv:0811.4622}} [hep-ph]].
  \relax
 \relax
\bibitem{Hoeche:2010kg}
S.~H{\"o}che, F.~Krauss, M.~Sch{\"o}nherr and F.~Siegert, \emph{{NLO matrix
  elements and truncated showers}}, JHEP \textbf{08} (2011),
  \href{http://www.slac.stanford.edu/spires/find/hep/www?eprint=arXiv:1009.112%
7}{123},  [\href{http://arXiv.org/pdf/1009.1127}{{\tt arXiv:1009.1127}}
  [hep-ph]]. \relax
 \relax
\bibitem{Hoeche:2012yf}
\href{http://inspirehep.net/record/1123387}{S.~H{\"o}che, F.~Krauss,
  M.~Sch{\"o}nherr and F.~Siegert}, \emph{{QCD matrix elements + parton
  showers: The NLO case}},  \href{http://arXiv.org/pdf/1207.5030}{{\tt
  arXiv:1207.5030}} [hep-ph]. \relax
 \relax
\bibitem{Gehrmann:2012yg}
\href{http://inspirehep.net/record/1123388}{T.~Gehrmann, S.~H{\"o}che,
  F.~Krauss, M.~Sch{\"o}nherr and F.~Siegert}, \emph{{NLO QCD matrix elements +
  parton showers in $e^+e^-\to$hadrons}},
  \href{http://arXiv.org/pdf/1207.5031}{{\tt arXiv:1207.5031}} [hep-ph]. \relax
 \relax
\bibitem{Aad:2012en}
G.~Aad et~al., ATLAS Collaboration collaboration, \emph{{Study of jets produced
  in association with a W boson in $pp$ collisions at $\sqrt{s} = 7$ TeV with
  the ATLAS detector}}, Phys.Rev. \textbf{D85} (2012),
  \href{http://inspirehep.net/record/1083318}{092002},
  [\href{http://arXiv.org/pdf/1201.1276}{{\tt arXiv:1201.1276}} [hep-ex]].
  \relax
 \relax
\bibitem{Aad:2011tqa}
G.~Aad et~al., ATLAS Collaboration collaboration, \emph{{Measurement of
  multi-jet cross sections in proton-proton collisions at a 7 TeV
  center-of-mass energy}}, Eur.Phys.J. \textbf{C71} (2011),
  \href{http://www.slac.stanford.edu/spires/find/hep/www?eprint=1107.2092}{176%
3},  [\href{http://arXiv.org/pdf/1107.2092}{{\tt arXiv:1107.2092}} [hep-ex]].
  \relax
 \relax
\bibitem{Chatrchyan:2011wn}
S.~Chatrchyan et~al., CMS Collaboration collaboration, \emph{{Measurement of
  the Ratio of the 3-jet to 2-jet Cross Sections in $pp$ Collisions at
  $\sqrt{s} = 7$ TeV}}, Phys.Lett. \textbf{B702} (2011),
  \href{http://www.slac.stanford.edu/spires/find/hep/www?eprint=1106.0647}{336%
--354},  [\href{http://arXiv.org/pdf/1106.0647}{{\tt arXiv:1106.0647}}
  [hep-ex]]. \relax
 \relax
\bibitem{Hoeche:2012ft}
\href{http://inspirehep.net/record/1086175}{S.~H{\"o}che, F.~Krauss,
  M.~Sch{\"o}nherr and F.~Siegert}, \emph{{W+n-jet predictions with MC@NLO in
  Sherpa}},  \href{http://arXiv.org/pdf/1201.5882}{{\tt arXiv:1201.5882}}
  [hep-ph]. \relax
 \relax
\bibitem{Hoeche:2012fm}
\href{http://inspirehep.net/record/1127523}{S.~H{\"o}che and M.~Sch{\"o}nherr},
  \emph{{Uncertainties in NLO + parton shower matched simulations of inclusive
  jet and dijet production}},  \href{http://arXiv.org/pdf/1208.2815}{{\tt
  arXiv:1208.2815}} [hep-ph]. \relax
 \relax
\end{thebibliography}

\end{document}